\documentclass[letterpaper,twocolumn,11pt]{article}
\usepackage{usenix,epsfig,endnotes}
\usepackage{times}
\usepackage{helvet}
\usepackage{courier}
\usepackage{color}
\usepackage{geometry}
\usepackage{array,booktabs}
\usepackage{amsmath}
\usepackage{verbatim}
\usepackage{graphicx}
\usepackage{url}

\usepackage{listings}
\newcommand\SA[1]{\textcolor{red}{Sandeep: #1}}

\usepackage{titling}
\begin{document}

\date{}

\title{\Large \bf StreetX: Spatio-Temporal Access Control Model for Data}

\author{
{\rm Sandeep Singh Sandha}\\
University of California, Los Angeles\\
Los Angeles, California 90095\\
sandha@cs.ucla.edu
}

\maketitle

\thispagestyle{empty}

\subsection*{Abstract}
Cities are a big source of spatio-temporal data
that is shared across entities to drive potential use cases. Many of the  
Spatio-temporal datasets are confidential and are selectively shared. To allow selective sharing, several access control models exist, however user cannot express arbitrary space and time constraints on data attributes using them.
In this paper we focus on spatio-temporal access control model.
We show that location and time attributes of data may decide its confidentiality via a motivating example and thus can affect user's access control policy.  In this paper, we present StreetX which enables user to represent constraints on multiple arbitrary space regions and time windows using a simple abstract language.  
StreetX is scalable and is designed to handle large amount of spatio-temporal data from multiple users.
Multiple space and time constraints can affect performance of the query and may also result in conflicts. StreetX 
automatically resolve conflicts and optimizes the query evaluation with access control to improve performance. We implemented and tested prototype of StreetX using space 
constraints 
by defining region having 1749 polygon coordinates on 10 million data records. Our testing shows that StreetX extends the current access control with spatio-temporal capabilities.

\subsection*{Acknowledgement}
This paper contains original work. All the figures, tables and diagrams were made by me. The paper is written by me. I have done implementation, evaluation and testing of the system. The project idea originated from my discussions with my advisor Prof. Mani B. Srivastava and Dr. Bharathan Balaji. They both helped me to formulate the problem and set goals to push the limit of current access control. Prof. Mani B. Srivastava also gave me valuable feedback to include data sharing policies.

I would like to acknowledge the help of Prof. Yuan Tian  in understanding the current access control models. 
Also, I would like to thank Dr. Bharathan Balaji, Amr Alanwar and Moustafa Alzantot for proofreading.


\section{Introduction}
Individuals, organizations and different sectors of city generate 
spatio-temporal data having volume, variety, velocity and value.
Few examples of 
such datasets include 
New York City taxi  data with over 1.1 billion taxi trips \cite{nyctaxidata:2017}, location tracking data 
of individuals,
 satellite data from NASA that produces 4TB of new data everyday \cite{nasadata:2017}, Los Angeles city traffic data \cite{LATrafficData:2017} and water data of 1.5 million sites \cite{USGSData:2017}. 
 These datasets has potential to 
create innovative applications
which encourage cities to collect more data. For example traffic and crash datasets are used together to predict safety of roads \cite{krumm2017risk}.
 To amplify usability, datasets are shared, stored and transformed by different entities. For instance, acute respiratory syndrome disease was controlled by the efforts of World Heath Organization only within 4 months after its emergence using extensive data sharing \cite{destro2014perspectives}.
 Several open data portals give public access to city information \cite{LAData:2017,USOpenData:2017}.
But some datasets are confidential and are only shared with authorized entities or shared partly. For example, the health department may share confidential data with law enforcement agencies. 

We 
focus on the space and time attributes 
of 
data to define access control policy.
Space and time attributes often decide the privacy and confidentiality of data. For example, let us consider Alice who is participating in a health study. During the study she monitors her multiple vitals and location for several days while doing normal daily activities. Bob is interested in health information of Alice for his research. Alice is willing to share her data so as to understand her health problems better. But Alice wants selective data sharing with Bob according to her privacy needs. She wants to share her data 
when she is 
within Los Angeles excluding her home  and only during working hours with a hourly time resolution. 
Alice also wants Bob to not share her data further. 
{In order to restrict access to data
different types of access control models exist, but they 
lack support to 
specify spatio-temporal constraints on data attributes. 
Thus, Alice cannot express her desired access control using earlier models. Alice may have to create and share subset of her data manually.}

 Access control based on space and time attributes 
restricts data at confidential locations and periods. Controlling the resolution of data limits the application usability and ability of a user to extract sensitive information. 
Data sharing policies affect the capabilities of user to share data further and thus can control data dissemination.
{However, providing spatio-temporal access control on large datasets is not trivial. Multiple arbitrary space and time constraints may result in conflicts and can also severely penalize the performance of the system. Thus it require conflict resolution strategies and policy optimization techniques.}

In this paper, we present StreetX which provides simple language to specify multiple arbitrary constraints on space and time attributes and data sharing policies. 
StreetX policy language present 
constraints
in human understandable semantics. { StreetX automatically resolves conflicting access control policies and uses optimization strategies when applying policies on user queries. 
 }
We show that using StreetX, Alice can easily express her access control policy for Bob. 
To evaluate the proposed access control model, we implemented and tested a prototype of StreetX to evaluate random user query along with space constraints specified using polygon of latitude and longitude having 1749 coordinates on 10 Million spatio-temporal data records. StreetX filters user queries using policy constraints and does query evaluation in 72 milliseconds on an average. 
 

{The remainder of this paper is organized as follows: in section 2  we discuss the related work; section 3 introduces the StreetX model; section 4 discusses access control and language. section 5 presents StreetX architecture and implementation details; section 6 gives evaluation and testing results; finally, in section 7 we conclude and point to future works.}

\section{Related Work}
{In this section, we survey access control models, policy definitions and related systems and illustrate that none of these support multiple arbitrary spatio-temporal constraints on data attributes.} In order to control the granularity of access control many techniques have been proposed. Traditional models such as Role-Based Access Control \cite{sandhu1996role} and  Attribute-Based Access Control \cite{hu2015attribute} are used extensively. Role-Based Access Control regulates access 
based on the role of individual user. User's role is associated with privileges and functions which user can perform. 
Attribute-Based Access Control uses attributes 
of user, data, context or action to restrict access. Multiple extensions to Role-Based Access Control using user's space and time context 
are proposed by researchers. For example TRBAC \cite{bertino2001trbac} 
activates user roles at certain time periods and LRBAC \cite{ray2006lrbac} 
controls user privileges based on her physical location. There are similar works 
\cite{le2012strobac,samuel2007framework,toahchoodee2009ensuring,ray2008spatio}
where user's role gets activated based on location or time of user requesting access to resource.  However they don't consider the space and time attributes of datasets to define spatio-temporal access control.
PlexC \cite{le2012plexc}
is a policy language designed for exposure control of data by allowing users to define access control policies based on different 
roles, disclosure level, time and location rules based on user context. PlexC doesn't consider multiple space regions and time windows on data attributes.

PDVLoc \cite{mun2014pdvloc} is designed for location based mobile services using which individual user can control location data sharing. PDVLoc is an individual data store with fine granular access control, policy recommendation and trace audit. PDVLoc supports space policies based on circular range. 
SensorSafe \cite{choi2011sensorsafe} also provides fine granular access control with major focus on privacy rules and data obfuscation techniques.  It allows single space region definition in access control. Both SensorSafe and PDVLoc have no support for multiple arbitrary space boundaries, time windows and data sharing policies.

Cell level access control also exists in databases like Apache Hbase \cite{hbase:2017} and Apache Accumulo \cite{accumulo:2017}. In Accumulo and Hbase each data cell can be associated with a label which can be later used to selectively control data access. Along with support for simple labels, logical \textit{AND} and \textit{OR} operations on labels are also supported. However these databases don't support spatio-temporal access control. ArcGIS \cite{arcgis:2017} is an industry standard geographic information system, but it only supports
Role-Based Access Control model. 
Data sharing policies are explored by attaching policies with data by Saroiu et al \cite{saroiu2015policy}.
Every time an organization request access to the user's data, it has to explicitly acknowledge the user policy however, policies are not enforce.  Organization may later misuse the data against the sharing policy which may have legal consequences. {We extend the idea to specify data sharing policies along with spatio-temporal access control. 
However, StreetX doesn't enforce the data sharing policies. 
}

{Collectively, these systems represent an effort to develop fine granular access control. However, no system allows multiple arbitrary constraints on space and time attributes of data in a multi-user setting which require expressible language model, conflict resolution and query optimization techniques. Also handling handling large datasets require distributed architecture to be scalable.}

\section{StreetX Model}
StreetX serves as a data hub for spatio-temporal data belonging to different entities. Streetx allows data owner to define fine granular spatio-temporal access control based on data attributes and specifying data sharing policies. 
Heterogeneous spatio-temporal datasets are transformed into a collection of data streams. 
Each data stream has an owner which specifies the access control policy and data sharing policy for other users. Both owners and users can query the data streams to which they have access to. StreetX supports rich spatio-temporal query across multiple data streams. While evaluating each query StreetX makes sure that the access control policy associated with the individual data stream is respected. Below we define the concept of data stream, owner, user and policy in our system.

\textbf{Data Stream}: Data Stream uniquely identifies the type of data. For example we may have two data streams one for the temperature sensor and other for the accelerometer sensor. A dataset with multiple measurements can be transformed into multiple data streams, one for each measurement. Each data stream consists of (latitude, longitude, time, value) and is uniquely identified by a data stream id ({$d_i$}). The set of data streams in the system is $D=\{ d_1,...,d_n\}$.

\textbf{Owner}: Every data stream has an owner. Owner is the entity who has created the data stream 
and has added data into it. Owner may own multiple data streams. Owner is the one which control the access and data sharing policies of the data stream. Owner creates policies for users with whom she wants to share her data stream. For simplicity we consider a data stream can have only one owner and ownership cannot be transfered. Every owner is uniquely identified by an owner id ({$o_i$}). The set of owners in the system is $O =\{o_1,...,o_m\}$.

\textbf{User}: User is the entity interested to use the data stream of owner. User can query the data stream, during which the owner policies defined on the data stream for the user are respected. A user may have access to several data streams and can query them simultaneously. In case user is the owner of the data stream, her query is answered directly. Every user is uniquely identified by a user id ({$u_i$}). The set of users in the system is $U =\{u_1,...,u_l\}$. {The set of owners ($O$) is a subset of the set of users ($U$).}

\textbf{Policy}: Policy defines the access of user to the data stream. 
There are two types of policies which we consider in StreetX: access control policy and data sharing policy. Access control policy is specified in terms of space and time attributes of the data stream. Access control policy can also affect the resolution of space and time attributes of data.
Data sharing policy defines the sharing capabilities of the user for the data.
Owner may define multiple policies on her data streams for same user. 
Every policy is uniquely identified by a policy id ({$p_i$}). The set of policies in the system is $P =\{p_1,...,p_r\}$.
\section{Policy \& Language}
StreetX 
has two different types of policies: Access control policies and data sharing policies. StreetX policy definitions are using abstract language constructs presented in Table  \ref{table:1}.
\begin{table}[h!]
\centering
\begin{tabular}{|c|c|} 
\hline
 \textbf{Constructs} & \textbf{Meaning \& Examples} \\ 
 \hline
 \textbf{What} & Specifies the set of data streams \\
 & on which policies are defined using \\ 
 &data  stream ids. Eg: What(\textbf{$d_1$, $d_2$})\\ 
\hline
 \textbf{Where} & Specifies the multiple semantic  \\
 &space regions  along with a  \\
 & optional NOT keyword to   \\
 &represent exclusion.\\
 &Eg: Where(\textbf{LA},  NOT  \textbf{HOME})\\
 

 \hline
 \textbf{When} & Specifies the multiple semantic \\
  &windows  of time  along  with a \\
  &optional  NOT  keyword to  \\
&represent exclusion. \\
&Eg: When(\textbf{WorkingHours},\\
& ``11/1/2016-11/31/2016")\\

 \hline
 \textbf{How} &  Specifies the resolution restrictions \\
 &on space and time. Resolution\\
 & keywords  are reserved. \\
 &Eg: How(\textbf{Hour}, \textbf{ZipCode})\\
 \hline
 \textbf{Whom} & Specifies the set of users to whom\\
 &this policy applies using user ids.\\
 &Eg: Whom(\textbf{$u_1$, $u_2$})\\
 
 \hline
 \textbf{Who} & Specifies the data sharing policy\\
 &of owner using reserved keywords.\\
 &Eg: Who(\textbf{DenyDataSharing})\\
 \hline
\end{tabular}
\caption{Policy Language}
\label{table:1}
\end{table}
In Table \ref{table:1}, the  \textit{What} construct select the set of data streams on which the policy is being applied and the \textit{Whom} construct selects the set of users whose access control is affected by the policy. Any policy at minimum must have  \textit{What} and \textit{Whom} constructs.
The \textit{Where}, \textit{When} and \textit{How} constructs are defining the access control policy based on space and time attributes of data, which is detailed in the 
Section \ref{AccConPol}.
The \textit{Who} construct defines the data sharing policy explained in Section \ref{datasharingpolicy}. In \textit{Where} construct, union and exclusion of different space regions can be defined and similar is allowed with time windows in \textit{When} construct.

\subsection{User Defined \& Language Keywords} \label{keywords}
Table \ref{table:1} shows a subset of keywords used in our language.
There are two types of keywords: user defined keywords and language reserved keywords. The user defined keywords are used in \textit{Where} and \textit{When} constructs to define space regions and time windows. For example the \textit{LA, HOME} and \textit{WorkingHours} are user defined keywords in Table \ref{table:1}. 
{We assume that in future, keywords can be defined using graphical user interface extension to StreetX, which allows  uploading GeoJSON \cite{GeoJSON:2017} definitions of regions and specify time constraints.} GeoJSON can be used to define arbitrarily complex space region. 
The language reserved keyword  \textit{NOT} can be used in \textit{Where} and \textit{When} constructs to exclude a particular space region or time window. The sample user keyword definitions are shown in Table \ref{table:2}.

\begin{table}[h!]
\centering
\begin{tabular}{|c|c|} 
\hline
 \textbf{Keywords} & \textbf{Definitions} \\ 
 \hline
 HOME
 & \{``Name":``HOME",``Type":\\
 &``Where",``Polygon":[\{$lat_1$,$lng_1$\}\\  
 &,\{$lat_2$,$lng_2$\},...,\{$lat_1$,$lng_1$\}]\}\\
\hline

WorkingHours
 & \{``Name":``WorkingHours",\\
 &``Type":``When",``RepeatedHour"\\
 &:``9AM-5PM",``ExcludeDay":\\
 &[``saturday",``sunday"]\} \\
\hline

\end{tabular}
\caption{User Defined Keywords}
\label{table:2}
\end{table}

The language reserved keywords to specify the resolution in \textbf{How} construct and data sharing policy specification in \textbf{Who} construct are shown in Table \ref{table:3}. 

\begin{table}[h!]
\centering
\begin{tabular}{|c|c|} 
\hline
 \textbf{Construct} & \textbf{Keywords} \\ 
 \hline
 How:Time
 & Second, Minute, Hour, Day,\\
 &  Week, Month, Year.\\
\hline

 How:Space
 & ZipCodes, County, City, Country.\\
\hline
 Who
 & AllowDataSharing, \\
 & DenyDataSharing, \\
 & PolicyUpdateEffect.\\
\hline
\end{tabular}
\caption{Language Keywords}
\label{table:3}
\end{table}

\subsection{Access Control Policy} \label{AccConPol}
Access control policy can be specified using the abstract language constructs which are shown in Table \ref{table:1}. The constructs are very near to the user's natural language and 
uses semantic keywords.  Looking back to our previous example of Alice and Bob, where Alice wanted to selectively share her historical health data with Bob. The access control policy to specify the space within Los Angeles excluding the home, during the working hours and with hourly resolution can be defined by Alice using Equation \ref{eq:1}, where $d_h$ identifies the health data stream of Alice, and $u_b$ identifies the user Bob. The \textit{HOME} and \textit{WorkingHour} keywords are defined in Table \ref{table:2}. Similarly \textit{LA} keyword can also be defined. 
 
\begin{equation} \label{eq:1}
\begin{split}
\text{What($d_h$).Where(LA, NOT HOME).}\\
\text{When(WorkingHours).How(Hour).Whom($u_b$)}
\end{split}
\end{equation}


\subsection{Data Sharing Policy}\label{datasharingpolicy}
Owner can specify data sharing policy 
in the \textit{Who} construct 
described in Table \ref{table:1} using the language reserved keywords defined in Table \ref{table:3}. \textit{Who(AllowDataSharing)} specifies that owner has allowed data sharing. The data shared by the user is at the resolution specified by the \textit{How} construct.  \textit{Who(AllowDataSharing,PolicyUpdateEffect)} specifies that data sharing is allowed by owner, but in future owner may change the policy to deny, and all the data shared by the user to multiple other entities will not be available to them. \textit{PolicyUpdateEffect} controls the effect of policy update by owner on multi-hop sharing. \textit{Who(DenyDataSharing)} specifies that data sharing is not allowed. 
{StreetX allows owner to specify data sharing policies which are informed to users. Violating these policies may have legal consequences. However StreetX doesn't enforces the data sharing policies.}
{The data sharing policies affect both raw data and processed data. We didn't separate this because user may use identity function on the raw data stream to replicate it or may encrypt the raw data stream values which might be used to regenerated raw data stream later.} In our previous example, the final policy of Alice for Bob along with specified data sharing is expressed in Equation \ref{eq:datasharing}. 

\begin{equation} \label{eq:datasharing}
\begin{split}
\text{What($d_h$).Where(LA, NOT HOME).}\\
\text{When(WorkingHours).How(Hour).Whom($u_b$)}\\
\text{.Who(DenyDataSharing)}
\end{split}
\end{equation}

\begin{figure}[h]
\includegraphics[width=0.5\textwidth]{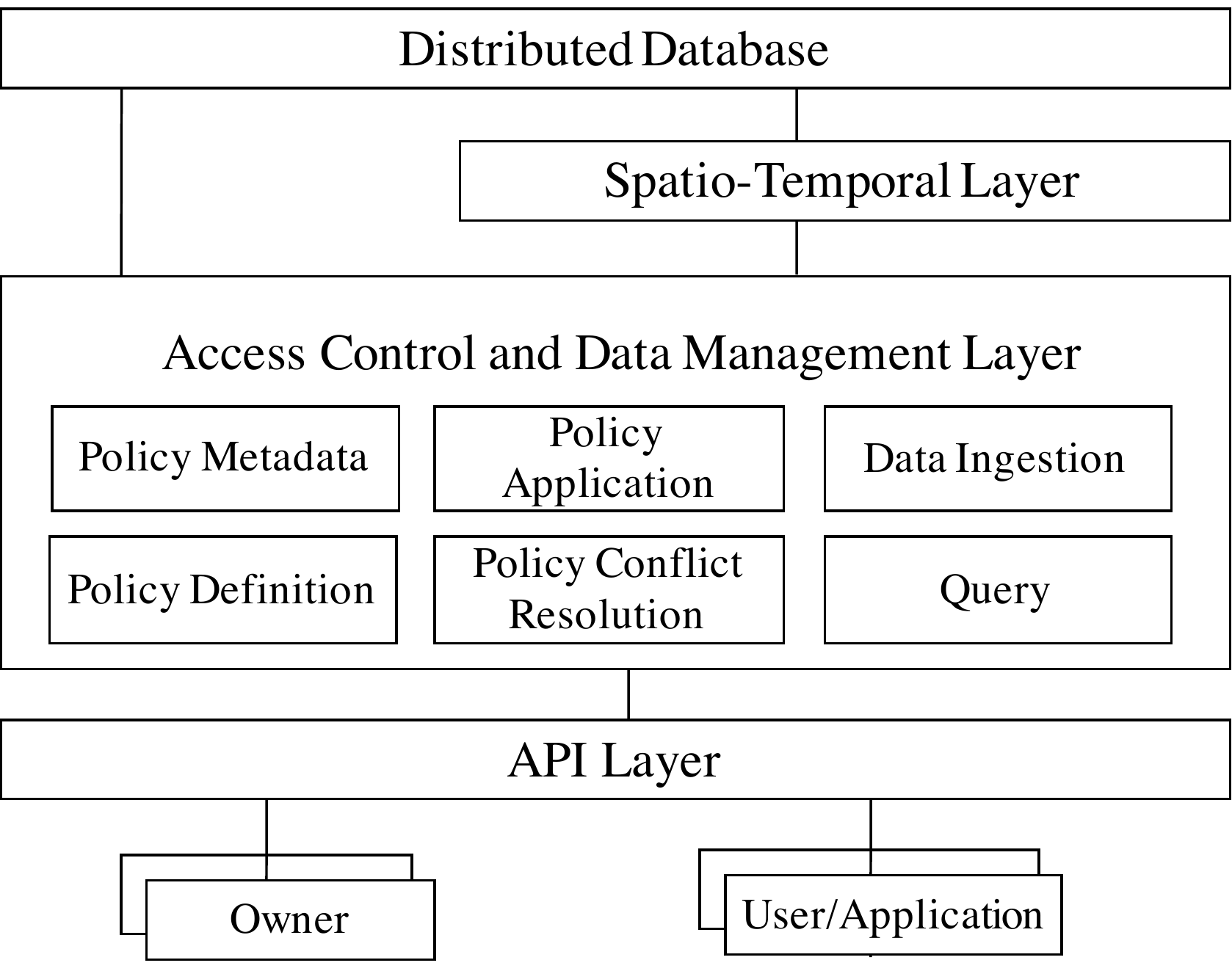}
\caption{System Architecture}
\label{fig:arch}
\end{figure}

\subsection{StreetX Workflow}
In order to understand the workflow of StreetX, let us consider our earlier example of Alice and Bob. Alice defines a data stream $d_h$ and uploads her entire health data.
She defines keywords mentioned in Table \ref{table:2} and creates a policy shown in Equation \ref{eq:1} 
for user Bob identified by $u_b$. Now, Bob can query Alice data, however data exposed to Bob is controlled by the policy of Alice. Our assumption is that health data of Alice is defined by only one data stream. She may define different data stream for each heath measurement type and can define access control policies on set of data stream. Suppose later Alice discovered that, her data within UCLA campus contains some confidential information, she can update her policy for Bob to hide her data while she is in UCLA campus and also keep the previous policy constrains as shown in Equation \ref{eq:workshop}. StreetX allows Alice to define and update her policy easily.

\begin{equation} \label{eq:workshop}
\begin{split}
\text{What($d_h$).Where(LA, NOT HOME, NOT}\\
\text{UCLA).When(WorkingHours).How(Hour)}\\
\text{.Whom($u_b$).Who(DenyDataSharing)}
\end{split}
\end{equation}

\section{StreetX Architecture \& Implementation}
StreetX architecture is shown in Figure \ref{fig:arch}. \textit{Distributed Database} is used for scalability and performance reasons. \textit{Distributed Database} stores the data streams, metadata defining the owners, users and policies. \textit{Spatio-Temporal} layer provides the capabilities which are used to enforce the access control policies based on space and time. \textit{Access Control and Data Management layer} have multiple components. \textit{Data Ingestion} module is used to ingest spatio-temporal data into the  \textit{Distributed Database}.  \textit{Policy Definition} module provides abstract language constructs presented in Table \ref{table:1} and keywords definition support discussed in section \ref{keywords}. \textit{Policy Metadata} module manages the information about policies, owners, data streams and users affected by policies. \textit{Query} module is used by owners and users to query the spatio-temporal data to which they have access to. \textit{Query} module uses information from \textit{Policy Metadata} module to identify the set of relevant policies for a particular query. The conflicts in the policies if any are resolved by the \textit{Policy Conflict Resolution} module. The policies are applied by \textit{Policy Application} module which translate the user query along with extra 
constraints introduced due to the policies into a spatio-temporal query. This transformed query is now evaluated by the \textit{Spatio-temporal layer}. The \textit{API Layer} exposes the functionality of StreetX  to authenticated owners and users. 

We implemented a prototype of StreetX in Java. We used Apache HBase \cite{hbase:2017} for \textit{Distributed Database} layer. Apache HBase is an open-source NoSQL scalable distributed database. Apache HBase stores the spatio-temporal data streams, and metadata.
We use GeoMesa \cite{geomesa:2017} as spatio-temporal layer on top of Apache HBase. GeoMesa is open source suite of tools that provide geospatial analytics support for spatio-temporal data. 
{GeoMesa supports Common Query Language \cite{cql:2017} 
which offers geospatial, temporal and attribute operators.}

\subsection{Data Ingestion \& Query}
\textit{Data Ingestion} module allows owner to push data into StreetX. StreetX supports the ingestion of spatio-temporal data expressed in the form of data streams. 
The data ingestion module inserts data into Apache HBase. \textit{Query} module exposes data streams to user. StreetX implementation supports spatio-temporal query on arbitrary set of data streams via JavaScript Object Notation (JSON) \cite{json:2017} format presented in Representation \ref{lst:querytrans}. 

{
\begin{lstlisting}[caption={User Query},captionpos=b,label=lst:querytrans]
{"(*\bfseries userId *)":$u_b$,"(*\bfseries DsID *)":[$d_1$,...,$d_i$],
"(*\bfseries SpaceBox *)":[$lat_{min}$,$lat_{max}$,$lng_{min}$,$lng_{max}$],
"(*\bfseries TimeRange *)":[$TimeStamp_{min}$,$TimeStamp_{max}$]}
\end{lstlisting}
}

\subsection{Policy Application}
\begin{figure}[h]
\centering
\includegraphics[width=0.5\textwidth]{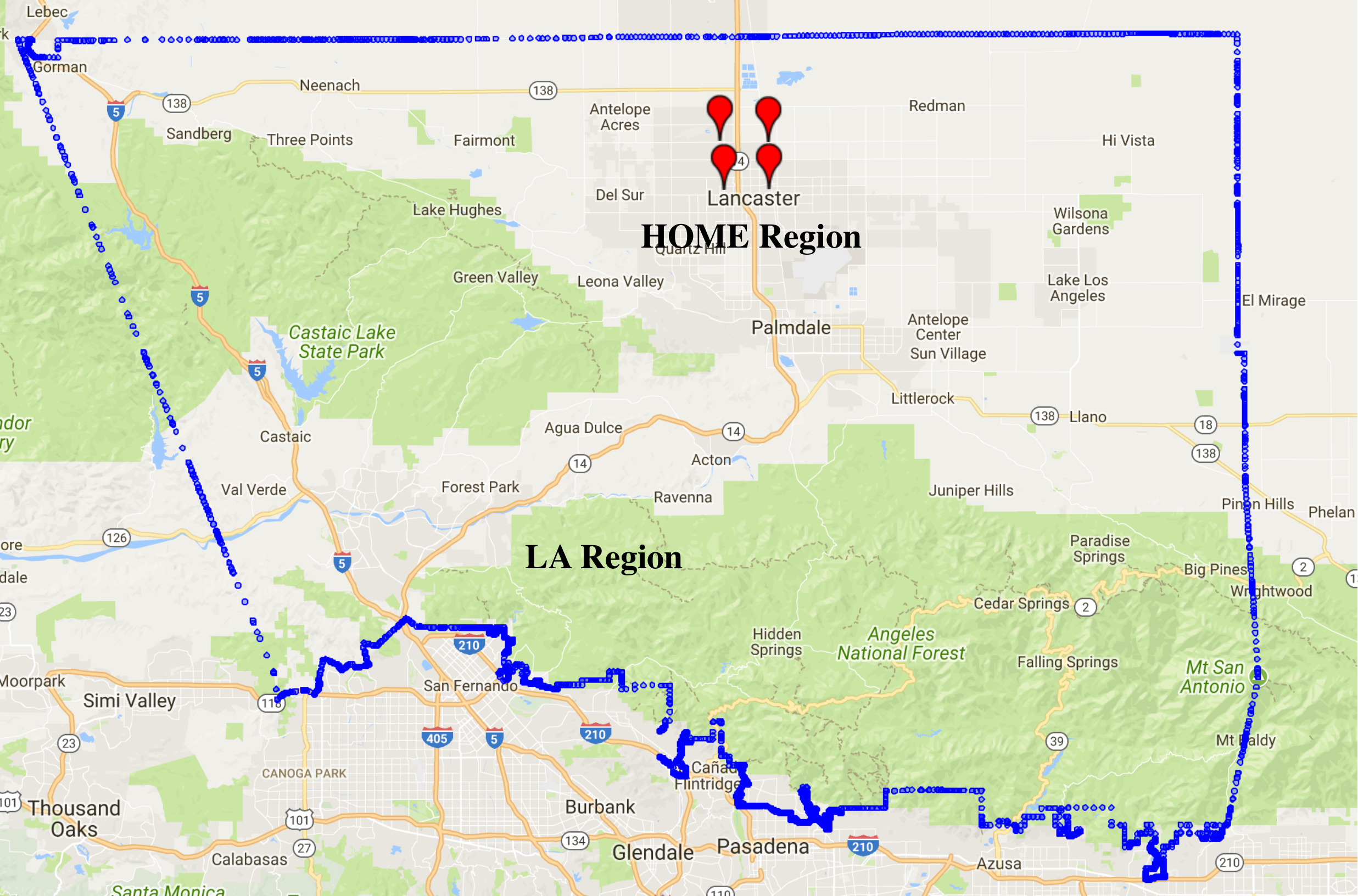}
\caption{ Space Region for Translation 1. Showing a subset of Los Angeles (\textit{LA}) Region having 1749 coordinates in blue and definition of \textit{HOME} Region using bounding box with 4 coordinates highlighted by red markers.}
\label{fig:trans1}
\end{figure}

\textit{Policy Application} module of Figure \ref{fig:arch} translates the policy into constraints on space and time attributes of data. \textit{Policy Application} module also performs the required computation for controlling resolution, which is discussed in Section \ref{resolution}. For example to apply policy of Alice specified in Equation \ref{eq:datasharing}, we translate the policy to respective space and time constraints in   JSON  using keyword definitions of Table \ref{table:2}. 
The sample output of translation of Equation \ref{eq:datasharing} is shown in Representation \ref{lst:Sptrans}. In order to translate \textit{WorkingHours}, we use the meta data of data stream $d_h$ to know the start time and end time of data and create time windows of daily  working hours in UNIX timestamps excluding Saturdays and Sundays. Visualizing the space region for this policy may look as shown in Figure \ref{fig:trans1}. 
In Figure \ref{fig:trans1}, 
 we have shown only 2 regions for simplicity, but policy language can express multiple arbitrary 
regions. 


{
\begin{lstlisting}[caption={Space and Time Constraints},captionpos=b,label=lst:Sptrans]
{"(*\bfseries Space *)":{"Allow":[{"Keyword":"(*\bfseries LA*)",
"Polygon":[{"lat":$lat_{i1}$,"lng":$lng_{i1}$},
..]}],"Deny":"[{"Keyword":"(*\bfseries HOME*)",
Polygon:[{"lat":$lat_{j1}$,"lng":$lng_{i1}$},
..]}]},
"(*\bfseries Time *)":[{"Keyword":"(*\bfseries WorkingHours*)",
Allow:[{"start:":$t_1$,"end":$t_2$},..]}]}
\end{lstlisting}
}

User query presented in Representation \ref{lst:querytrans} and policy translation presented in Representation \ref{lst:Sptrans} are further translated into the effective set of constraints for evaluation. The constraints can be grouped into set of space polygons and set of time windows for each data stream. For simplicity if we consider a single data stream, the final effective constraint translation is presented in Representation \ref{lst:finaltrans}. Here \{$R_{qb}$\}$_{QuerySpaceBOX}$ and \{$T_{1}$,$T_{2}$\}$_{QueryTimeRange}$ are from user query (here Bob). \{$R_{LA}$\}$_{SpacePolicy}$, \{$R_{HOME}$\}$_{SpacePolicy}$ and \{$T_{i-1}$,$T_{i}$\}$_{QueryTimeRange}$ are from access control policy of Alice given in Equation \ref{eq:1}. 
In prototype implementation, StreetX allows the policy constraints expressed in Representation 2. We plan to extend it with the proposed language constructs in future.
Representation \ref{lst:finaltrans} can be directly translated into Common Query Language supported by by GeoMesa, however it may adversely affect the performance.
We use query optimization strategies discussed in Section \ref{queryoptimization} to improve the query evaluation time.

{
\begin{lstlisting}[caption={Example Final Effective Constraints},captionpos=b,label=lst:finaltrans]
{"(*\bfseries Space *)":{{$R_{qb}$}$_{QuerySpaceBOX}$ AND
{{$R_{LA}$}$_{SpacePolicy}$ AND {NOT $R_{HOME}$}$_{SpacePolicy}$}},
"(*\bfseries Time *)":[{$T_{min}$,$T_{max}$}$_{QueryTimeRange}$ AND 
{{$T_{1}$,$T_{2}$}$_{QueryTimeRange}$,... AND {$T_{n-1}$,$T_{n}$}}]
\end{lstlisting}
}

\subsection{Query Optimization} \label{queryoptimization}
{ Final access control constraints shown in Representation \ref{lst:finaltrans} can affect the query evaluation performance. In our testing discussed in Section \ref{testing}, considering the direct user query on data stream without access control as baseline. The performance is degraded by a factor of 4 times due to access control constraints of Figure \ref{fig:trans1} even though less number of data points are fetched. 
We reject the user query and return 0 results if it cannot be satisfied by looking at all the constraints. In order to do this, we check the satisfiability of Representation \ref{lst:finaltrans} to see if user query intersects the allowed space regions and time windows specified by policy. For example, to checking the satisfiability of spatial constraints in Representation \ref{lst:finaltrans}, we do the spatial query to see if \{$R_{qb}$\}$_{QuerySpaceBOX}$  intersects \{$R_{LA}$\}$_{SpacePolicy}$ and is not completely within \{$R_{HOME}$\}$_{SpacePolicy}$. StreetX keeps policy constraints in memory and applies them to every query before translating it and evaluating using GeoMesa. If, we directly evaluate an unsatisfiable query on GeoMesa, it will still return 0 results, but takes significantly more time which affects query performance as discussed in Section \ref{testing}.  
In future, We also plan to group together the similar constraints because complexity of contraints directly affect the performance as discussed in Section \ref{testing}. For example, two space regions can be grouped together, if one of them is completely within the other.
} 


\subsection{Space \& Time Resolution} \label{resolution}
StreetX handles resolution constraints on a particular data stream by creating a separate data stream with the required resolution. For example, to implement the hourly resolution, we convert all the time values within an hour to the same value, making them indistinguishable in time attribute, the similar concept is extended to space attribute. For spatial resolution, we assume the availability of required spatial boundaries. StreetX redirects all the user queries to new data stream 
 which is not exposed to owner directly and exists only if resolution policy is defined by the Owner. 
 Other approach is to create resolution on the fly from original data stream after doing the query. It will require processing of time and location values for every record in the result set and may also discard some of the records, because resolution can also affect the number of points fetched. We didn't use this approach due to performance reason, however replicating data stream uses extra storage. 
In prototype implementation, StreetX assumes the data stream with desired resolution. 

\subsection{Policy Conflict Resolution}
\begin{figure}[h]
\includegraphics[width=3in]{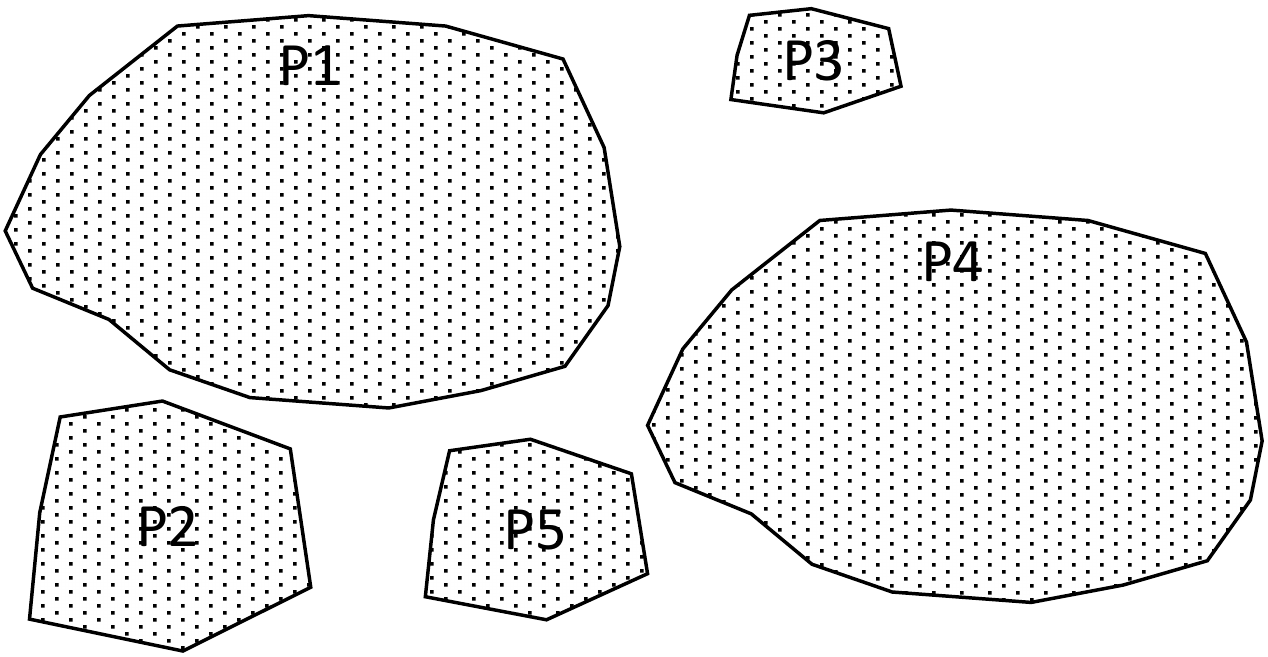}
\caption{Non-Overlapping Space Regions for Different Policies. }
\label{fig:trans2}
\end{figure}

\textit{Policy Conflict Resolution} module identifies conflicting policies at the time of policy definition. 
Allowing multiple space and time restrictions may result in conflicts. 
Conflict happens when multiple policies exist for the same user defined over the same data stream. 
Three different scenario's in conflict are possible: Firstly, When either the space regions or time windows from different policies don't overlap with each other. Such a scenario is explained in Figure \ref{fig:trans2}, where policy P1, P2, P3, P4 and P5 have space region constraints of type allow. In this case, union of all policies is followed , even though their time windows may overlap. In such conflicts, multiple policies are defined over different subsets of data within data stream and thus are not really conflicting in terms of constraints, a union of policies is followed by doing union on all the constraints. 

\begin{figure}[h]
\includegraphics[width=3in]{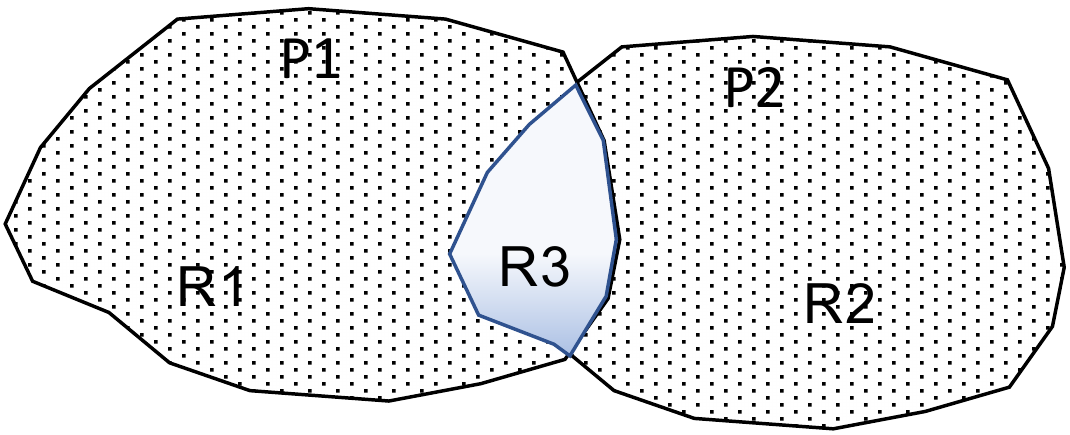}
\caption{Overlapping Space Region for Policy P1 and Policy P2.}
\label{fig:trans3}
\end{figure}

Second scenario happens when both space regions and time windows of different policies 
overlap, which means both are defined over same set of data within data stream. Consider for example two policies P1 and P2 where space region and time windows of both policies overlap. For simplicity, consider space regions are as shown in Figure \ref{fig:trans3} and time windows are exactly same for both policies. {In this case, StreetX follows the principle of least privilege while doing union of policies. StreetX gives high priority to the constraints denying space regions and time windows.
}

A third scenario which requires special treatment is an extension to second scenario. It happens when policies have  different resolutions. For example, consider the Figure \ref{fig:trans3}. If policy P1 and policy P2 have different resolution, then enforcing resolution constraints is a challenge.  Currently, StreetX create different data streams for policy P1 and Policy P2, if they have different resolution. This will require StreetX to subdivide the query and use resolution of P1 for R1 and resolution of P2 for R2. But for R3, decision cannot be simply made based on principle of least privilege because two different resolutions may not be comparable sometime. For example, resolution based on time cannot be compared with resolution based on space. 
StreetX model cannot resolve the third scenario completely 
without attaching priorities with policies. We have not implemented \textit{Policy Conflict Resolution} module in prototype of StreetX. We assume that all the desired constraints are already grouped together by following union of policies and are expressed in Representation 2. 

\section{Evaluation and Testing}\label{testing}





We deployed prototype of StreetX on a single instance of Microsoft Azure \cite{azure:2017} DS4\_V2 having 8 cores and 28GB RAM on Ubuntu Server 16.04 using 
HBase v1.2.5.   
We tested the performance of StreetX in applying space policy constraint using a region shown in Figure 2, having 7649 coordinates extracted from the website of Los Angeles County Department of Public Works \cite{LAcountyshape:2017}. The policy definition 
used in this testing is shown in Equation \ref{eq:streetxtesting}.

\begin{equation} \label{eq:streetxtesting}
\begin{split}
\text{What($d$).Where(LA, NOT HOME).Whom($u$)}\\
\end{split}
\end{equation}

The \textit{LA} space region has a complex shape as shown in Figure \ref{fig:trans1} which is roughly 5,000 $km^2$.
Table \ref{table:3} presents summary of testing.  A data stream having 
10 Million points was generated randomly with location variation in an area $A$ of 500 KM by 500 KM around Los Angeles 
and time variation $T$ of 1 year between 2014 to 2015. 
User query is generated randomly in the format shown in Representation \ref{lst:querytrans} using a \textit{spacebox} of 50KM by 50KM within $A$ and \textit{timerange} of 2 weeks within $T$.
Three different types of queries are performed as shown in Table \ref{table:3}. The results presented are the average of 1000  user queries. ${Query_d}$ is direct user query without any policy application. It gives the baseline for number of points fetched with no space restrictions and performance time.  ${Query_a}$ is user query evaluated along with policy of Equation \ref{eq:streetxtesting} without optimizations discussed in Section \ref{queryoptimization}. ${Query_b}$ is user query evaluated along with policy of Equation \ref{eq:streetxtesting} and optimizations discussed in Section \ref{queryoptimization}. 

\begin{table}[h]
\centering
\begin{tabular}{|c|c|c|c|} 
\hline
 \textbf{Dataset} & \textbf{Query} &
 \textbf{No. of points} &
 \textbf{Time (ms)} \\
 \hline
 10 Million
 &$Query_d$ & 5230 &540\\
\hline

 10 Million
 &  $Query_a$  &202 &2325\\
\hline
 10 Million
 & $Query_b$ & 202 &72\\
\hline
\end{tabular}
\caption{StreetX Testing}
\label{table:4}
\end{table}
$Query_d$ serves as a baseline when user query is evaluated without any policy.  As expected with policy restrictions the number of points fetched decreased from 5230 to 202 in $Query_a$ and $Query_b$. But $Query_a$ takes 4 times more time then even $Query_d$ though it retrieves less points. This extra time is due to the complex space policy restrictions.  $Query_b$ takes only 72 milliseconds to evaluate on an average. The reason behind this is rejection of the queries which cannot be satisfied by \textit{Policy Application} module without evaluating them using GeoMesa. In our experiments of generating 1000 random user queries, 900 queries were rejected by \textit{Policy Application} module using optimizations discussed in Section \ref{queryoptimization}.  $Query_b$ also keeps the policies in memory which does the language translation faster. In order to understand the effect of number of coordinates in space constraint, we performed $Query_a$ and $Query_b$ by decreasing the number of coordinates in \textit{LA} space region from 7649 by 3 times to 2550  by random selection. The results are shown in Table \ref{table:5}, which depict the effect of size of constraints on performance. The results indicate that user should express the resolution of space boundary with the least possible number of coordinates. The increase in performance of $Query_a$ is more than 6 times whereas for $Query_b$ it is less, which is because $Query_b$ is already optimized.  Our results show that StreetX can express and evaluate complex spatio-temporal access control policies over large datasets.

\begin{table}[h]
\centering
\begin{tabular}{|c|c|c|c|} 
\hline
 \textbf{Dataset} & \textbf{Query} &
 \textbf{No. of points} &
 \textbf{Time (ms)} \\
\hline

 10 Million
 &  $Query_a$  &258 &345\\
\hline
 10 Million
 & $Query_b$ & 258 &63\\
\hline
\end{tabular}
\caption{StreetX Testing With 2550 Space Constraint Coordinates}
\label{table:5}
\end{table}

\section{Conclusions \& Future Works}
In this paper, we presented StreetX to enhance access control capabilities  with fine granular constraints on space and time attributes of data. 
StreetX serves novel applications by allowing selective data sharing based on confidential locations and time periods using simple abstract language. 
One such evaluation was presented by us where space policy was defined using complex space region. Arbitrary spatio-temporal constraints may result in conflicts which are handled by StreetX. 
StreetX also uses query optimization strategies which enhance performance significantly. Our evaluation shows StreetX is scalable and can represent and evaluate arbitrary spatio-temporal policies over large datasets in an optimized manner.

In future, we plan to implement all the proposed features and do further testing of StreetX by evaluating it using real spatio-temporal datasets. We plan to integrate StreetX with city scale data hub projects like MetroInsight \cite{MetroInsight:2017} and health data hub initiatives like MD2K \cite{MD2K:2017}  to enhance data sharing of real systems. Different extensions to StreetX are possible, like allowing users to specify policies in natural language 
and built in support for general shape 
definitions of cities, counties etc. We also plan to explore use of trusted computing platforms \cite{pearson2002trusted} to enforce data sharing policies.

\footnotesize \bibliographystyle{acm}
\bibliography{main}

\end{document}